\documentclass[twocolumn,showpacs,preprintnumbers,amsmath,amssymb]{revtex4}
\usepackage{tipa}
\usepackage{amssymb}
\usepackage{stmaryrd}
\usepackage{txfonts}
\usepackage{graphicx}% Include figure files
\usepackage{dcolumn}% Align table columns on decimal point
\usepackage{bm}% bold math

\begin{document}

\preprint{}

\title{Giant quantized Goos-H\"{a}nchen effect on the surface of graphene in quantum Hall regime}
% Force line breaks with \\
\author{Weijie Wu}
%\altaffiliation[ ]{}%Lines break automatically or can be forced with \\
\author{Shizhen Chen}
\author{Chengquan Mi}
\author{Wenshuai Zhang}
\author{Hailu Luo}\email{hailuluo@hnu.edu.cn}
\author{Shuangchun Wen}
%%\email{Second.Author@institution.edu}
\affiliation{Laboratory for Spin Photonics, School of Physics
and Electronics, Hunan University, Changsha 410082, China}
\date{\today}% It is always \today, today,

\begin{abstract}
We theoretically predict a giant quantized Goos-H\"{a}nchen (GH) effect on the surface of graphene in quantum Hall regime.
The giant quantized GH effect manifests itself as an angular shift whose quantized step reaches the order of
mrad for light beams impinging on a graphene-on-substrate system. The quantized GH effect can be attributed to quantized Hall conductivity, which corresponds to the discrete Landau levels in quantum Hall regime.
We find that the quantized step can be greatly enhanced for incident angles near the Brewster angle.
Moreover, the Brewster angle is sensitive to the Hall conductivity, and therefore the quantized GH effect
can be modulated by the Fermi energy and the external magnetic field.
The giant quantized GH effect offers a convenient way to determine
the quantized Hall conductivity and the discrete Landau levels by a direct optical
measurement.
\end{abstract}


%Use showkeys class option if keyword
\maketitle

\section{Introduction}\label{SecI}
The well-known Snell's law and Fresnel formulae provide a clear geometrical-optics
picture to describe the interaction of a plane wave with an interface~\cite{Jackson1999,Born2005}.
In 1947, the spatial shift of a beam of light in total internal reflection
from a dielectric surface was demonstrated which dose not follow perfectly the geometric optics prediction.
This spatial shift was firstly observed by F. Goos
and H. H\"{a}nchen~\cite{Goos1946}, and was therefore referred to as the Goos-H\"{a}nchen (GH) effect.
In a simple explanation, such a spatial GH shift is attributed to the penetration of evanescent field.
For the past few decades, the spatial GH shift has been studied in a variety
of systems, such as plasmonics~\cite{Yin2004,Yin2006,Salasnich2012}, metamaterials~\cite{Wild1982,Pfleghaar1993,Emile1995,Bonnet2001,Berman2002,Felbacq2003,Shadrivov2003,Felbacq2004,He2006,Longhi2011,Grosche2016,Xu2016},
and quantum systems~\cite{Beenakker2009,Lee2014}.
In addition, the angular GH shift has been predicted
for the case of partial reflection, which can be explained as the Fresnel filtering~\cite{Tureci2002,Schomerus2006}. This
remarkable deviation from geometric optics has been
measured experimentally on the surface of bulk crystals~\cite{Merano2009,Merano2010,Araujo2016,Araujo2017}.

Recently, graphene, as a two-dimensional atomic crystal,
has received considerable attention due to its
extraordinary electronic and photonic properties~\cite{Novoselov2004,Neto2009,Bonaccorso2010}.
It has been demonstrated that the Fresnel formulae
based on the certain thickness and effective refractive index
fails to perfectly explain the the light-matter
interaction of graphene. However, the Fresnel formulae based on the
zero-thickness interface can give a complete and convincing
description of all the experimental observations~\cite{Merano2016II,Chen2017}.
It would be interesting how the GH effect occurs on the zero-thickness interface of graphene.
More recently, the quantized spatial shifts in GH effect
have been theoretically predicted in the quantum Hall regime of graphene-substrate systems~\cite{Kamp2016}.
However, the quantized steps are just a fraction of a micrometer in terahertz regime.
Therefore, how to enhance this tiny effect is still a challengeable problem.

In this paper, we theoretically predict a giant quantized GH effect on the surface of graphene in quantum Hall regime.
A general propagation model
is established to describe the GH
shifts on the surface graphene-on-substrate system and freestanding graphene.
Based on this model, both the spatial and the angular GH
shifts are obtained when a light beam impinges on the surface of graphene.
Most of previous works have demonstrated that the beam shifts can be significantly enhanced near Brewster angle
on the surface of bulk crystals~\cite{Chan1985,Chang2009,Qin2009,Kong2012,Zhou2012I,Zhou2012II,Pan2013,Zhou2013III,Gotte2014,Grosche2015,Hermosa2016}.
As excepted, a giant angular GH shifts are also obtained on the surface graphene.
More importantly, we find that the quantized steps in angular GH shifts can be significantly enhanced and reaches the order of
mrad. Furthermore, we examine the role of the Hall conductivity in quantized GH effect.
We believe this work to be fundamental significance and maybe provide a possible scheme for the direct
optical measurement of the quantized effect in graphene.

\section{General propagation model}\label{SecII}
\begin{figure}
\centerline{\includegraphics[width=8.5 cm]{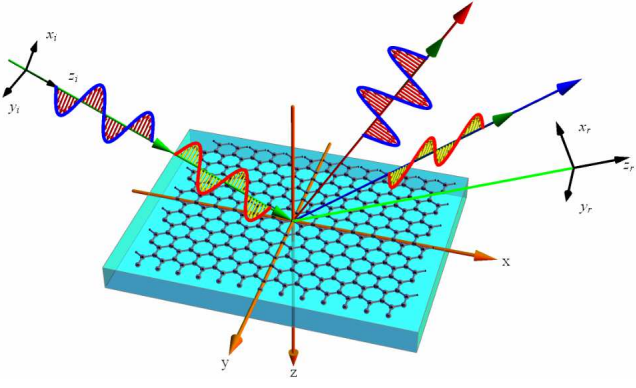}}
\caption{\label{Fig1} Schematic representation of the
wave reflection at a graphene-substrate interface. The homogeneous and isotropic substrate
is covered by graphene sheet. An external imposed
static magnetic field $B$ is applied perpendicular
to the interface. On the reflecting surface,
the giant angular GH shifts occur different values, which the shift of
parallel polarized reflection is greater than perpendicular
polarized reflection.}
\end{figure}

We begin by analyzing optical reflection from a
planar interface of graphene-substrate system.
The Fig.~\ref{Fig1} illustrates a monochromatic Gaussian
beam of light with finite beam width and
non-totally reflection impinging from air to a planar
interface of graphene-substrate system. The $z$ axis of
the laboratory Cartesian frame ($x,y,z$) is normal to
the air-graphene interface ($z=0$), that
separates empty space (typically air), where $z<0$,
from a substrate that is covered with a graphene sheet,
where $z>0$, and a static magnetic field $B$ is applied
along the $z$ axis. We use the coordinate frames ($x_i,y_i,z_i$)
and ($x_{r},y_{r},z_{r}$)
to denote incident beam and reflected one, respectively.
The electric field amplitude of such a beam can be
written as~\cite{Li2007,Aiello2008,Bliokh2013}
\begin{equation}
\tilde{E}_{i}\propto\exp\left[ikz_i-\frac{k}{2}\frac{x_{i}^2+y_{i}^2}{Z_R+iz_i}\right]\times(\hat{\textbf{x}}_{i}f_{p}+\hat{\textbf{y}}_{i}f_{s})\label{equ1},
\end{equation}
where $Z_R=\pi w_0^2/\lambda$ is the Rayleigh range, the vectors
$\mathbf{\hat{x}}, \mathbf{\hat{y}}$ represent the directions of parallel
and perpendicular to the incidence plane, respectively. And the
polarization of the beam is determined by the complex-valued
unit vector $\mathbf{\hat{f}} = (f_p \mathbf{\hat{x}}_i+f_s \mathbf{\hat{y}}_i)/(|f_p|^2+|f_s|^2)^{1/2}$.

The reflected angular spectrum of the electric field is associated
with the boundary distribution by means of the relation~\cite{Luo2011}
\begin{eqnarray}
\left[\begin{array}{cc}
\tilde{E}_r^p\\
\tilde{E}_r^s
\end{array}\right]
=  \left[
  \begin{array}{cc}
    r_{pp} & r_{ps} \\
    r_{sp} & r_{ss} \\
  \end{array}
\right]\cdot\left[\begin{array}{cc}
\tilde{E}_i^p\\
\tilde{E}_i^s
\end{array}\right]\label{equ2}.
\end{eqnarray}
Here, $r_{pp}$ and $r_{ss}$ denote the Fresnel reflection coefficients
for parallel and perpendicular polarizations, respectively.
$r_{ps}$ and $r_{sp}$ denote cross-polarization.

In the Eq.~(\ref{equ2}), we have introduced the boundary condition
$k_{rx}=-k_{ix}$ and $k_{ry}= k_{iy}$. By making use of Taylor
series expansion based on the arbitrary angular spectrum component,
$r_{A}$ can be expanded as a polynomial of $k_{ix}$:
\begin{eqnarray}
r_{A}(k_{ix})&=&r_{A}(k_{ix}=0)+k_{ix}\left[\frac{\partial
r_{A}(k_{ix})}{\partial
k_{ix}}\right]_{k_{ix}=0}\nonumber\\
&&+\sum_{j=2}^{N}\frac{k_{ix}^N}{j!}\left[\frac{\partial^j
r_{A}(k_{ix})}{\partial k_{ix}^j}\right]_{k_{ix}=0}\label{Tarloy},
\end{eqnarray}
where $A\in\{pp,ss,ps,sp\}$.
Then, the reflected field can be solved by utilizing the
Fourier transformations. The complex amplitude
can be conveniently expressed as
\begin{eqnarray}
\mathbf{E}_r(x_r,y_r,z_r )&=&\int\int d k_{rx}dk_{ry}
\tilde{\mathbf{E}_r}(k_{rx},k_{ry})\nonumber\\
&&\times\exp [i(k_{rx}x_r+k_{ry}y_r+ k_{rz} z_r)],\label{equ5}
\end{eqnarray}
where $k_{rz}=\sqrt{k_r^2-(k_{rx}^2+k_{ry}^2)}$ and
$\tilde{\mathbf{E}_r}(k_{rx},k_{ry})$ is the reflected angular
spectrum.

From Eq.~(\ref{equ1}) - Eq.~(\ref{equ5}), the general expression
of the reflected field is determined and can be written as
\begin{eqnarray}
\mathbf{E}_{r}&\propto&
\exp\left(ikz_{r}-\frac{k}{2}\frac{x_{r}^2+y_{r}^2}{Z_R+iz_r}\right)\nonumber\\
&&\times\bigg\{\mathbf{\hat{x}}_r\bigg[f_{p}r_{pp}\bigg(1-\frac{ix_r}{Z_{R}+iz_r}\frac{\partial\ln r_{pp}}{\partial\theta_i}\bigg)\nonumber\\
&&+f_{s}r_{ps}\bigg(1-\frac{ix_r}{Z_{R}+iz_r}\frac{\partial\ln r_{ps}}{\partial\theta_i}\bigg)\bigg]\nonumber\\
&&+\mathbf{\hat{y}}_r\bigg[f_{s}r_{ss}\bigg(1-\frac{ix_r}{Z_{R}+iz_r}\frac{\partial\ln r_{ss}}{\partial\theta_i}\bigg)\nonumber\\
&&+f_{p}r_{sp}\bigg(1-\frac{ix_r}{Z_{R}+iz_r}\frac{\partial\ln r_{ps}}{\partial\theta_i}\bigg)\bigg]\bigg\}\label{equ6},
\end{eqnarray}
where $f_{p}=a_{p}\in\mathbb{R}$, $f_{s}=a_{s}\exp(i\eta)$.

In addition, the Fresnel reflection coefficients of the graphene-substrate
system with an external imposed magnetic field can be
obtained as~\cite{Tse2011,Kamp2015,Cai2017}
\begin{equation}
 r_{pp}=\frac{\alpha^T_+\alpha_-^L+\beta}{\alpha_+^T\alpha_+^L+\beta}\label{RPP},
  \end{equation}
  \begin{equation}
 r_{ss}=-\frac{\alpha^T_-\alpha_+^L+\beta}{\alpha^T_+\alpha_+^L+\beta}\label{RSS},
  \end{equation}
  \begin{equation}
r_{ps}=r_{sp}=-2\sqrt{\frac{\mu_0}{\varepsilon_0}}\frac{k_{iz}k_{tz}\sigma_H}{\alpha^T_+\alpha_+^L+\beta}\label{RPS}.
  \end{equation}
Here, $\alpha^L_\pm=(k_{iz}\varepsilon\pm k_{tz}\varepsilon_0+k_{iz}k_{tz}\sigma_L/\omega)/\varepsilon_0$,
$\alpha^T_\pm=k_{tz}\pm k_{iz}+\omega\mu_0\sigma_T$,
$\beta=\mu_0k_{iz}k_{tz}\sigma^2_H/\varepsilon_0$,
$k_{iz}=k_i\cos\theta_i$, and $k_{tz}=k_t \cos\theta_t$;
$\theta_t$ is the refraction angle; $\varepsilon_0$ , $\mu_0$  are
permittivity and permeability in vacuum, respectively; $\varepsilon$ is the permittivity of substrate;
$\sigma_L$, $\sigma_T$ and $\sigma_H$
denote the longitudinal, transverse, and Hall conductivity, respectively.

When the external imposed magnetic field is strong enough,
the Hall conductivity of the graphene is quantized in integer
multiples of the fine structure constant, and we have~\cite{Kamp2016}
\begin{equation}
\sigma_H=2(2n_c+1)\mathrm{Sgn}[B]\frac{e^2}{2\pi\hbar}\label{HallC}.
\end{equation}
Here, $n_c=\mathrm{Int}[\mu^2_F/2\hbar e|B|v^2_F]$
is the number of occupied Landau levels, $\mu_F$, and $v_F$ are the Fermi energy and the Fermi velocity, respectively.
Obviously, Landau levels play an important role in Hall conductivity.
Note that in the linear optical response of an important $2$D atomic crystal model
of Fresnel coefficients in graphene has been developed by fixing both the surface susceptibility and the
surface conductivity~\cite{Merano2016I}.

\section{the giant angular Goos-H\"{a}nchen Shifts}\label{SecIII}
In this section, we begin to reveal the giant angular GH shifts
in graphene and discuss the relation between the shift and incident
angle. Then we try to explore the relationship between the magnitude of
Brewster angle and external conditions of magnetic field
and Fermi energy.
We now determine the centroid of the reflected beam. At any given plane $z_r$ = const, the longitudinal displacement of field
centroid is given by
\begin{equation}
D_{GH} = \frac{\int \int x_r I(x_r,y_r,z_r)\text{d}x_r \text{d}y_r}{\int \int I(x_r,y_r,z_r)\text{d}x_r \text{d}y_r}.\label{centroid}
\end{equation}
The beam intensity spatial profile is closely linked to flux of the time averaged Poynting vector $I(x_r,y_r,z_r)\propto \bar{S}\cdot \hat{z}_r$.
Then the Poynting vector related to the electromagnetic
field can be obtained by $\bar{S}\propto \mathrm{Re}(\mathbf{E}_{r}\times\mathbf{H}_{r}^*)$.
The magnetic field can be obtained by $\mathbf{H}_r=-ik_{r}^{-1}\nabla\times \mathbf{E}_r$.

\begin{figure}
\centerline{\includegraphics[width=8.5 cm]{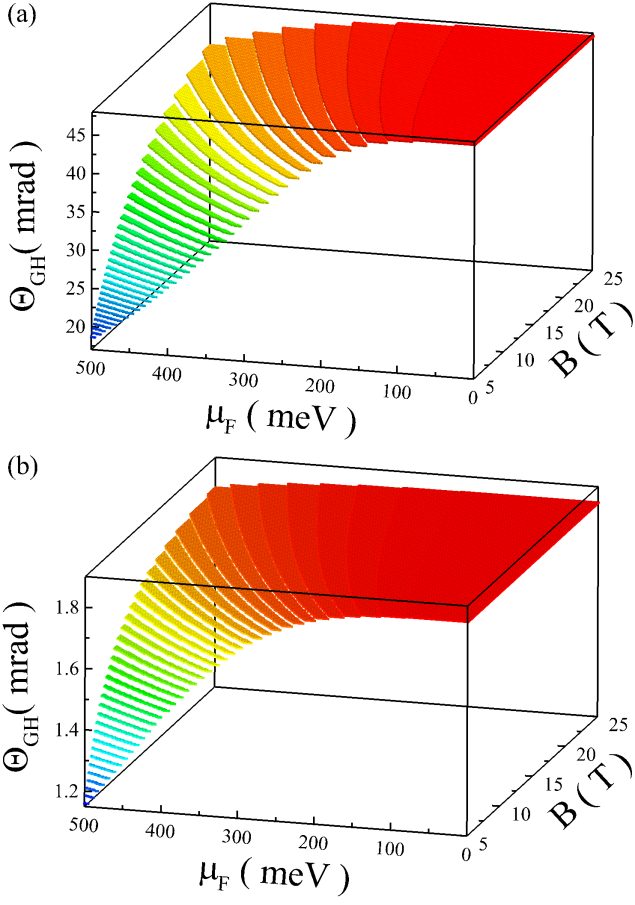}}
\caption{\label{Fig2} Quantized angular Goos-H\"{a}nchen shifts
in graphene-substrate system as
a function of Fermi energy and magnetic field.
(a) The incident angles are chosen as to $71$ degrees, which is near the Brewster angle,
and (b) $30$ degrees, which is faraway from the Brewter angle, respectively.
We assume incident beam with $w_0=1$ mm, $\omega/2\pi =1$ THz. The refractive index of undoped Si
in the terahertz range is $n_{Si}=3.415$.
The temperature is chosen as $T=4\mathrm{K}$.}
\end{figure}

In order to simplify the calculation,
the horizontal polarization is merely considered,
i.e. $a_{p}=1$, $a_{s}=0$, and $\eta=0$.
From  Eq.~(\ref{centroid}), we get the following expression
\begin{eqnarray}
  D_{GH} &=& \frac{2(R_{pp}^2\varphi_{pp}+R_{ps}^2\varphi_{ps})Z_{R}}{2k(R_{ps}^2+R_{pp}^2)Z_{R}+\chi_{pp}+\chi_{ps}} \nonumber\\
  &&-z_r\frac{2(R_{pp}^2\rho_{pp}+R_{ps}^2\rho_{ps})}{2k(R_{ps}^2+R_{pp}^2)Z_{R}+\chi_{pp}+\chi_{ps}}.\label{equ11}
\end{eqnarray}
Here, $r_A=R_{A}\exp(i\phi_A)$, $A\in\{pp,ss,ps\}$,
$\rho_{A}=$Re$(\partial\ln r_{A}/\partial\theta_{i})$,
$\varphi_{A}=$Im$(\partial\ln r_{A}/\partial\theta_{i})$, and $\chi_{A}=R_{A}^2(\varphi_{A}^2+\rho_{A}^2)$.
If we consider vertical polarization (i.e. $a_{p}=0$, $a_{s}=1$, and $\eta=0$),
we can replace the $pp$ with $ss$ in above eqution.
Furthermore, since the vertical polarization is considered,
the induced cross-polarization is $r_{sp}$ instead of $r_{ps}$.

Equation~(\ref{equ11}) gives the GH shift as a function of the beam propagation distance $z_r$.
The first term is considered to represent the spatial GH shift,
that is, the displacement will not change with $z_r$. If we use it in the conditions of total
internal reflection and isotropy, we could get a result that is consistent with the Artmann formula~\cite{Artmann1948}. Namely,
when we make $r_{ps}=0$ and $|R_{pp}|=1$, we will get $D_{GH}=(\partial\phi_{A}/\partial\theta_{i})/k$.
Then the second term denote the angular GH shift. For more general cases,
the derivative of the Fresnel reflection coefficients can be easily simplified,
so an equation can be obtained by $\partial\ln r_{A}/\partial\theta_{i}=(\partial R_{A}/\partial\theta_{i})/R_{A}+i\partial\phi_{A}/\partial\theta_{i}$.
That is, the change of phase and amplitude reflectivity is responsible for spatial and angle shifts, respectively.
We mainly discuss the angular shift, that is second term in above equation,
so we get the important result
\begin{equation}\label{equ12}
  \Theta_{GH} =-\frac{2(R_{pp}^2\rho_{pp}+R_{ps}^2\rho_{ps})}{2k(R_{ps}^2+R_{pp}^2)Z_{R}+\chi_{pp}+\chi_{ps}}.
\end{equation}

As shown in the Fig.~\ref{Fig2},
the angular GH shifts are quantized functions of the Fermi enengy and magnetic field,
Plateaulike behavior can be observed by tuning Fermi enengy, $\mu_{F}$, and magnetic field, $B$.
The quantized Hall conductivity can be regarded as the physical origin of Plateaulike behavior.
The quantized steps reach the order of mrad near the Brewster angle and can be determined
by a direct optical measurement~\cite{Santana2016}.
And, due to that the Hall conductivity is quantified, the quantized angular shift can be obtained.
We now consider the difference of angular deviation for incidence angle near and far away from Brewster angle[Fig.~\ref{Fig2}(a) and Fig.~\ref{Fig2}(b)].
Remarkably, the angular shift for incidence angle near the Brewster angle will be greater,
at the same Fermi energy and magnetic field.

\begin{figure}
\centerline{\includegraphics[width=8.5 cm]{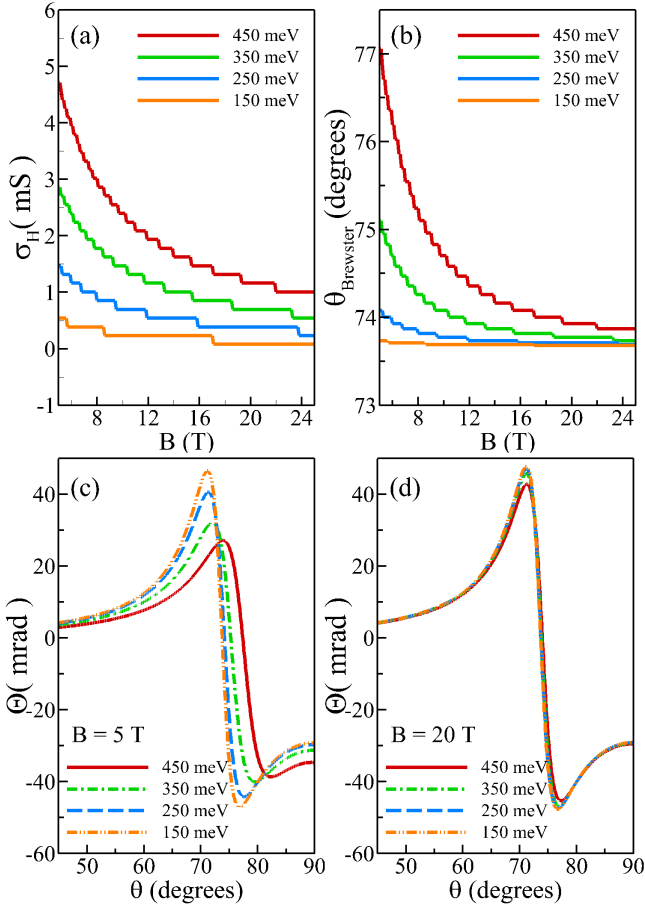}}
\caption{\label{Fig3} The role of quantized Hall conductivity in quantized angular GH shifts.
(a) Hall conductivity as function of magnetic field
in different Fermi energy, $\mu_{F}=150$, $250$, $350$, and $450$ meV.
(b) The magnitudes of Brewster angle.
(c) and (d) show the changes
of the angular shift in diffrent Fermi
energy and the magnetic field.}
\end{figure}

But if we research it further, we will find that angular shift is not only
influenced by incident angle,
the shift is also related to the quantized Hall conductivity. In fact,
the Landau levels are proportional to Fermi energy squared,
but they are inversely proportional to magnetic field.
And, Landau levels play an important role in Hall conductivity.
From Fig.~\ref{Fig3}(a), it can be seen that the Hall conductivity is decreased
as the Fermi energy decreased or the magnetic field increased.
But it is worthy noting that the quantized steps width can be significantly enhanced.
From Eq.~(\ref{RPP}),
the quantized steps width of Hall conductivity have a close relationship with Brewster angle.
In other words, from Fig.~\ref{Fig3}(b),
the Brewster angle will dramatically change in the case of narrow quantized steps.
This directly leads to the change of angular deviation.
In the region of high magnetic field, this phenomenon is not obvious.
This is consistent with our previous statement.
Due to the widen quantized steps, the Brewster angle is insensitive to the changing magnetic field and Fermi energy.
Compare Fig.~\ref{Fig3}(c) with Fig.~\ref{Fig3}(d), it can be seen that only when we maintain the
magnetic field at a relatively small value ($B=5$T), which leads a narrow quantized steps,
the peak of angular shift will become sensitive and move to the right
with the increase of the Fermi energy.
Of course, due to the magneto-optical effect,
the behavior of angular deviation in the graphene-substrate system is different from
that in the glass-air system~\cite{Merano2010}.
In addtion, the reflection coefficient, $r_{pp}$,
will approach zero near the Brewster angle and change its sign
across the angle, which means the electric field reverses its directions.
So, when the incident angle is smaller than the Brewster angle,
the angular shift is positive, and in the opposite case, the angular shift is negative.
For vertical polarization,
the angular deviation is very small.
This rule of change seems to be an enlightenment for
us, if we appropriately control the external conditions,
the angular shift can be modulated.

\begin{figure}
\centerline{\includegraphics[width=8.5 cm]{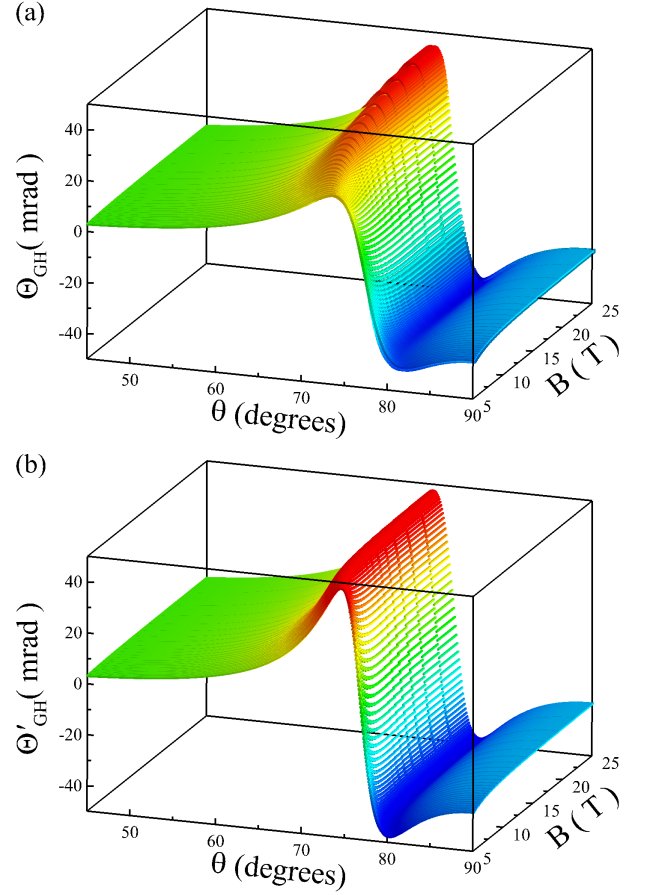}}
\caption{\label{Fig4} Compare the original and the modified angular GH shifts. (a) The giant quantized GH shift
in $450$ meV Fermi energy (similarly hereinafter).
(b) The modified angular GH shift.}
\end{figure}

Then, we consider two special cases. Firstly, if we used a polarizer
to eliminate cross-polarization component, namely $R_{ps}$,
or used a isotropic reflected medium, which has also not cross
polarization component $R_{ps}$ and $R_{sp}$ in reflection matrix
in the Eq.~(\ref{equ2}),
a modified angular GH shifts can be obtained. Here,
the horizontal polarization is only discussed.
We now consider the effect of cross-polarization component $R_{ps}$
on the GH shifts at different magnetic field.
So, we have modified expression
\begin{equation}\label{equ13}
  \Theta_{GH}'=-\frac{2R_{pp}^2\rho_{pp}}{2kR_{pp}^{2}Z_{R}+\chi_{pp}}.
\end{equation}

\begin{figure}
\centerline{\includegraphics[width=8.5 cm]{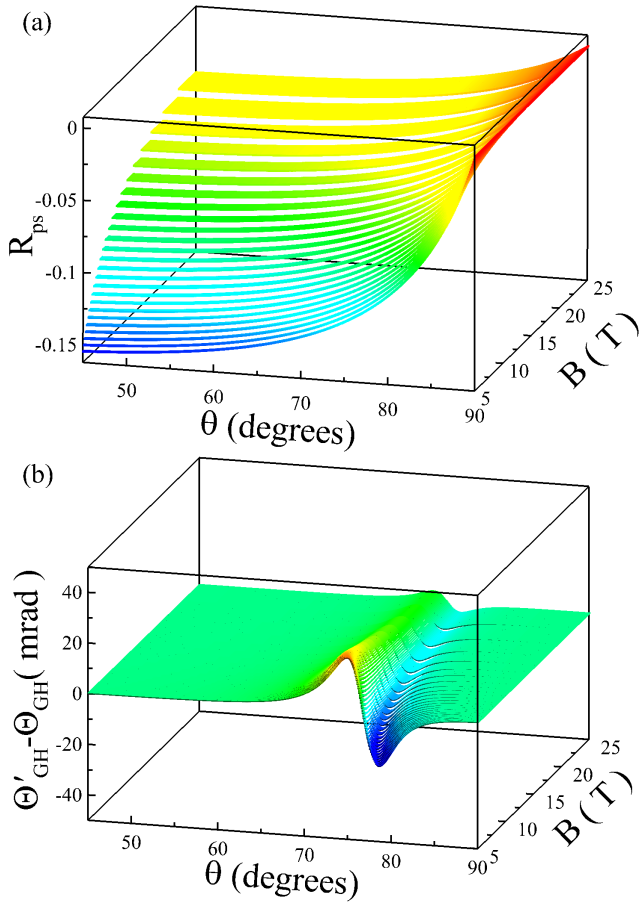}}
\caption{\label{Fig5} The role of cross-polarization in angular GH effect. (a) Relative amplitude of the cross-polarization, $R_{ps}$ ,
as a function of angle of incidence, $\theta$, and magnetic field, $B$, for Fermi energy $\mu_{F}=450$ meV.
(b) The difference between the normal GH shift, $\Theta_{GH}$, and the modified GH shift, $\Theta'_{GH}$.}
\end{figure}

\begin{figure}
\centerline{\includegraphics[width=8.5 cm]{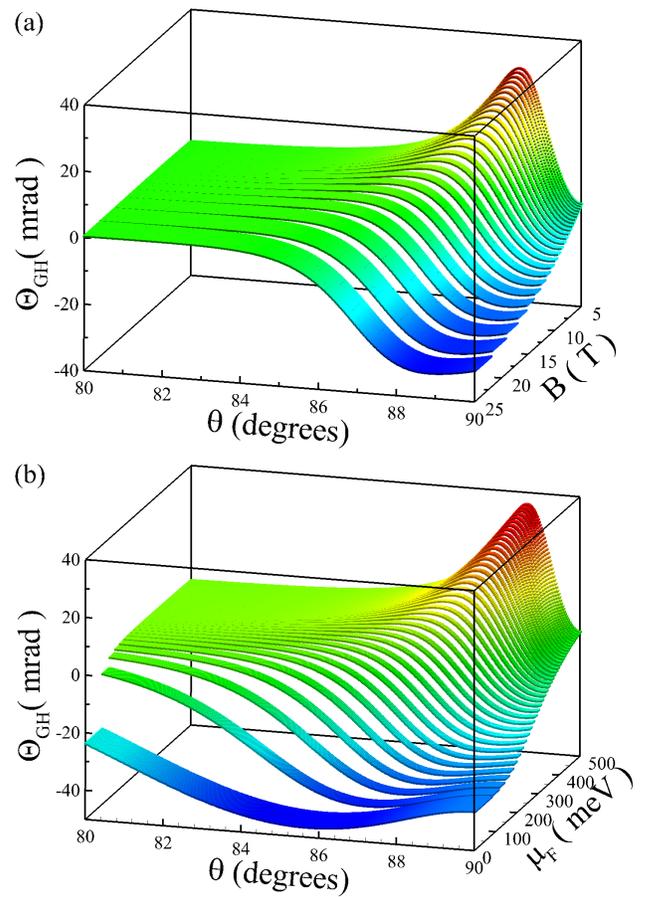}}
\caption{\label{Fig6} Angular GH shift in freestanding graphene.
(a) Angular GH shifts, $\Theta_{GH}$, as a function of incidence angle, $\theta$,
and magnetic field, $B$, for Fermi energy $\mu_{F}=450$ meV.
(b) Angular GH shifts, $\Theta_{GH}$, as a function of incidence angle, $\theta$,
and Fermi energy, $\mu_{F}$, for magnetic field $B=5$ T.
The relative refractive index of the substrate tend to $n=1$,
other parameters are the same as in Fig.~\ref{Fig2}.}
\end{figure}

Next, we compare the angular GH shift with the modified GH shift.
From Fig.~\ref{Fig4}(a), there are giant quantized angular GH shifts with the change of magnetic field.
And the peak of $\Theta_{GH}$
near the Brewster angle has an radian, which is more obvious in a area with narrow quantized steps.
This proves our previous statement: If quantized steps were narrowed,
the angular GH shifts would be sensitive to changing magnetic field and Fermi energy.
So, the position of peak will be moved.
From Fig.~\ref{Fig4}(b), the peak value of $\Theta_{GH}'$ near the area,
where magnetic field is $5$ T,
is approximately $20$ mrad larger than that of $\Theta_{GH}$ in the corresponding range,
that is, beam propagation will produce a deviation of $2$ centimeter per meter.

For further analysis of this difference,
it is plotted that the magnitude of the cross-polarization reflection coefficients,
$R_{ps}$. As shown in the Fig.~\ref{Fig5}(a), when magnetic field is decreased,
the magnitude of $R_{ps}$ will increase. In fact,
our previous statement still works.
Widen quantized steps will cause $R_{ps}$ to be insensitive to the changing magnetic field and Fermi energy.
From Fig.~\ref{Fig5}(b),
it is clear that a significant difference between the two case near the Brewster angle.
However, it is worthy to notice that, the difference will
decrease in the range far away from the Brewster angle or in the interval of widen quantized steps. That is to say,
there are no difference (or very small difference) of angular GH shifts in the above two range.
At this point, we could get
$\Theta_{GH}''=-\rho_{p}/kZ_R$, which is in good agreement with the
theoretical result of Aiello~\cite{Aiello2008}.
That is, the $R_{ps}$ is responsible for
difference between two shifts only near the Brewster angle or in the interval of narrow quantized steps.

Finally, we consider the case of freestanding graphene~\cite{Liu2017},
where we can make the relative refractive index
of the substrate being tend to $n=1$.
As shown in Fig.~\ref{Fig6},
it can be seen that when the refractive index approaches $1$,
the magnitude of incidence angle, which is corresponding to peak, is increased, that is,
the Brewster angle will approach grazing incidence on freestanding graphene.
On the other hand,
when the magnetic field is high or Fermi energy is low in quantum Hall regime,
the magnitude of Brewster angle is decreased,
the quantized steps are also wide at this time,
which proves that the Brewster angle in this moment is insensitive to changing magnetic field or Fermi energy,
namely, wide quantized steps lead to GH shift to be insensitive to changing magnetic field and Fermi energy.

\section{Conclusions}

In conclusion, we have theoretically predicted a giant quantized Goos-H\"{a}nchen (GH) effect on the surface of graphene in quantum Hall regime.
A strict model has been established and revealed a giant quantized angular GH shifts,
which is dominated by a change of reflectance,
for incidence angle near the Brewster angle on reflection.
The quantized steps of angular deviation have been greatly enhanced for incident angles near the Brewster angle.
We have found that when
magnetic field is high or Fermi energy is low in quantum Hall regime,
the quantized steps of Hall conductivity can be significantly widen.
Meanwhile, quantized Hall conductivity is related to the discrete Landau levels in quantum Hall regime.
In addition, we have demonstrated that cross-polarization component can not be ignored for incidence angle near Brewster angle or in the case of narrow quantized steps.
And we have found that Brewster angle would tend to grazing incidence in the case of freestanding graphene.
We can determine the quantized Hall conductivity and the discrete Laudau levels by a direct optical measurement.
These findings provide a pathway for modulating the GH effect and
thereby open the possibility of developing new nanophotonic devices.

\section*{ACKNOWLEDGMENTS}
This research was supported by the National Natural Science
Foundation of China (Grant No. 11474089).

\end{document}